\documentclass[twocolumn,
	prd,
	amssymb,
	preprintnumbers,
	secnumarabic,
	nofootinbib]{revtex4-1}
	
\pdfoutput=1

\usepackage{graphicx}
\usepackage{enumitem}
\usepackage{latexsym}
\usepackage{amsfonts}
\usepackage{amssymb}
\usepackage{color}
\usepackage{xcolor}
\usepackage{amsmath}
\usepackage{slashed}
\usepackage{dcolumn}
\usepackage{verbatim}
\usepackage{float}
\usepackage{multirow}
\usepackage[colorinlistoftodos]{todonotes}
\usepackage{xspace}
\usepackage[normalem]{ulem}
\usepackage[
pdfauthor={Manish Tamta}]{hyperref}

\newcommand{\gsim}{\gtrsim}
\newcommand{\lsim}{\lesssim}

\def\Oc{\mathcal{O}}

\renewcommand{\tilde}{\widetilde} 
\newcommand{\beq}{\begin{equation}}
\newcommand{\eeq}{\end{equation}}
\newcommand{\bea}{\begin{eqnarray}}
\newcommand{\eea}{\end{eqnarray}}
\newcommand{\nn}{\nonumber}

\definecolor{aquamarine}{rgb}{0.2,0.7,0.6}
\definecolor{cerulean}{RGB}{0,166,214} 
\definecolor{subtlered}{rgb}{0.8,0.3,0.3}

\hypersetup{
 pdfauthor={Manish Tamta, Nirmal Raj, Prateek Sharma},
  colorlinks=true,
  citecolor=cerulean,
  urlcolor=aquamarine,
  linkcolor=cerulean
}

\newcommand{\DLensSource}{D_{\rm LS}}
\newcommand{\DLens}{D_{\rm L}}
\newcommand{\DSource}{D_{\rm S}}
\newcommand{\RSource}{R_{\rm S}}
\newcommand{\aSource}{a_{\rm S}}
\newcommand{\Rschw}{R_{\rm Schw}}
\newcommand{\mubarE}{\bar \mu_{\rm E}}
\newcommand{\muthresh}{\bar \mu_{\rm thresh}}

\newcommand{\rE}{r_{\rm E}}
\newcommand{\tE}{t_{\rm E}}
\newcommand{\vE}{v_{\rm E}}

\newcommand{\rhodm}{\rho_{\rm DM}}

\allowdisplaybreaks

\usepackage{scalerel,tikz}
\usetikzlibrary{svg.path}
\definecolor{orcidlogocol}{HTML}{A6CE39}
\tikzset{orcidlogo/.pic={
 \fill[orcidlogocol] svg{M256,128c0,70.7-57.3,128-128,128C57.3,256,0,198.7,0,128C0,57.3,57.3,0,128,0C198.7,0,256,57.3,256,128z};
 \fill[white] svg{M86.3,186.2H70.9V79.1h15.4v48.4V186.2z}
 svg{M108.9,79.1h41.6c39.6,0,57,28.3,57,53.6c0,27.5-21.5,53.6-56.8,53.6h-41.8V79.1z M124.3,172.4h24.5c34.9,0,42.9-26.5,42.9-39.7c0-21.5-13.7-39.7-43.7-39.7h-23.7V172.4z}
 svg{M88.7,56.8c0,5.5-4.5,10.1-10.1,10.1c-5.6,0-10.1-4.6-10.1-10.1c0-5.6,4.5-10.1,10.1-10.1C84.2,46.7,88.7,51.3,88.7,56.8z};
}}
\newcommand\orcidicon[1]{\href{https://orcid.org/#1}{\mbox{\scalerel*{
\begin{tikzpicture}[yscale=-1,transform shape]
\pic{orcidlogo};
\end{tikzpicture}
}{|}}}}

\begin{document}

\title{Breaking into the window \\ of primordial black hole dark matter with x-ray microlensing}

\author{Manish Tamta$^{\orcidicon{0009-0006-7088-8705}}$} \email{manishtamta@iisc.ac.in}
\author{Nirmal Raj$^{\orcidicon{0000-0002-4378-1201}}$} \email{nraj@iisc.ac.in}
\author{Prateek Sharma$^{\orcidicon{0000-0003-2635-4643}}$} \email{prateek@iisc.ac.in}
\affiliation{Indian Institute of Science, C. V. Raman Avenue, Bengaluru 560012, India}

\date{\today}

\begin{abstract}
Primordial black holes (PBHs) in the mass range $10^{-16}-10^{-11}~M_\odot$ may constitute all the dark matter.
We show that gravitational microlensing of bright x-ray pulsars provide the most robust and immediately implementable opportunity to uncover PBH dark matter in this mass window.
As proofs of concept, we show that the currently operational NICER telescope can probe this window near $10^{-14}~M_\odot$ with just two months of exposure on the x-ray pulsar SMC-X1, and that the forthcoming STROBE-X telescope can probe complementary regions in only a few weeks.
These times are comparable to the week-long exposures obtained by NICER on various individual sources. 
We take into account the effects of wave optics and the finite extent of the source, which become important for the mass range of our PBHs.
We also provide a spectral diagnostic to distinguish microlensing from transient background events and to broadly mark the PBH mass if true microlensing events are observed.
In light of the powerful science case, i.e., the imminent discovery of dark matter searchable over multiple decades of PBH masses with achievable exposures, we strongly urge the commission of a dedicated large broadband telescope for x-ray microlensing.
We derive the microlensing reach of such a telescope by assuming sensitivities of detector components of proposed missions, and find that with hard x-ray pulsar sources PBH masses down to a few $10^{-17}~M_\odot$ can be probed. 
\end{abstract}

\maketitle

%%%%%%%%%%%%%%%%%%%%%%%%%
\section{Introduction}
\label{sec:intro}
%%%%%%%%%%%%%%%%%%%%%%%%%

\begin{figure}[h]
    \centering
    \includegraphics[width=0.49\textwidth]{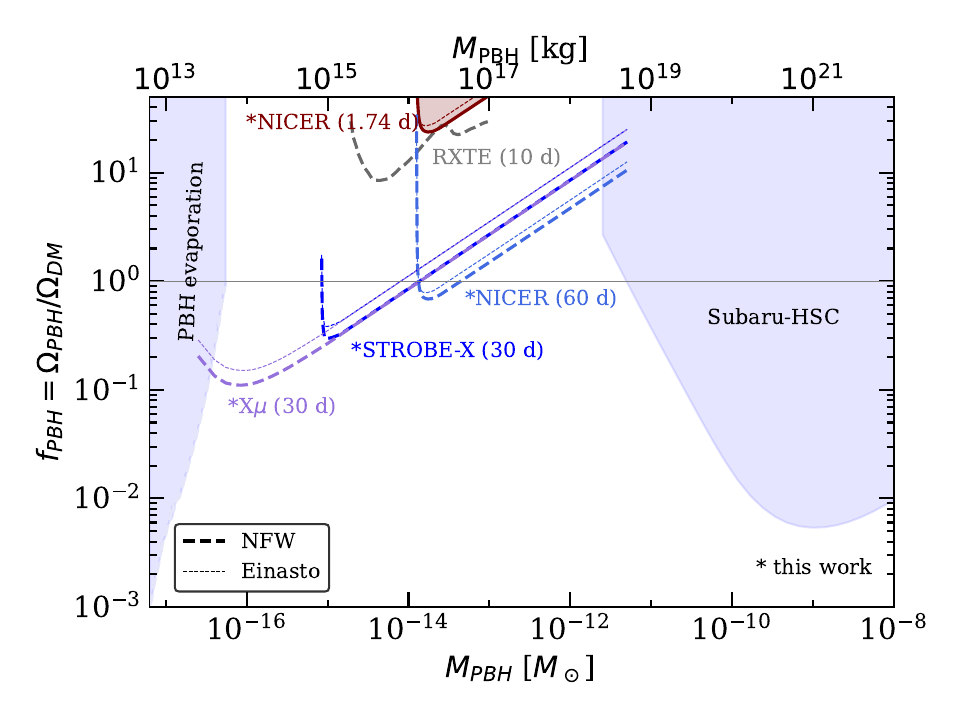}
    \caption{90\% C.L. constraints on primordial black hole dark matter from observations of the x-ray pulsar SMC X-1 over 1.74 days at NICER, and projected reaches of NICER (60 days on SMC-X1), the future STROBE-X (30 days on SMC-X1), and our proposed X$\mu$ (30 days on a Crab-like hard x-ray pulsar in the SMC).
    With one year of exposure at NICER, 
    $f_{\rm PBH}$ about six times smaller than the 60-day reach shown here can be constrained.
    Also shown are x-ray microlensing limits using RXTE~\cite{baiorlofsky}, optical microlensing using Subaru-HSC~\cite{CroonMcKeenRajECOlocation2}, and PBH evaporation~\cite{kavanaghgreenreviewPBH:2020jor}. 
    See Sec.~\ref{subsec:results} for further details.
    }
    \label{fig:fvM}
\end{figure}

One appealing contender for the make-up of the unidentified dark matter is primordial black holes (PBHs), formed likely due to density perturbations generated during the inflationary epoch of the universe~\cite{hawkingcarr1974,*KhlopovreviewPBH:2008qy,*evap:Carr:2009jm,*CarrreviewPBH:2016drx,*CarrreviewPBH:2020gox,*CarrreviewPBH:2020xqk,kavanaghgreenreviewPBH:2020jor}.
Hunted extensively by looking for signatures of their evaporation, gravitational microlensing, accretion-induced distortion of the relic radiation, gravitational waves from binary mergers, and dynamical effects in stellar systems, they have left open a mass window between about $10^{-16}-10^{-11}~M_\odot$ over which they can constitute 100\% of the dark matter; see Ref.~\cite{kavanaghgreenreviewPBH:2020jor} for the most up-to-date constraints. 
In this paper we show that gravitational microlensing performed at current and imminent x-ray telescopes can, for the very first time, realistically probe this mass window.
Our main result is shown in Figure~\ref{fig:fvM}, which we will discuss in detail in Sec.~\ref{subsec:results}.

%%%%%
%%%%

Microlensing is the temporary magnification of a background star due to the gravitational field of a transiting body, with images typically unresolved.
It is a technique that, using stellar sources in the Milky Way and neighbouring galaxies, has ruled out massive compact halo objects (MACHOs) and PBHs in the mass range $10^{-11}-10~M_\odot$ as the entire content of dark matter~\cite{CroonMcKeenRajECOlocation2}. 
The technique can also be applied, albeit in a non-trivial way, to extended objects with various density distributions~\cite{WidrowHClouds,Fairbairn:2017dmf,Blinov_2020,CroonMcKeenRajECOlocation2,CroonMcKeenRajECOlocation1,BaiDMACHOs:2020jfm,AnsariArunQBalls:2023cay}.
The smallest PBH masses reached by microlensing were constrained by the Subaru-HSC instrument~\cite{Niikura2019Subaru,Smyth:2019whb,CroonMcKeenRajECOlocation2}, which was sensitive to the short transit times of light PBHs.
However, the Subaru-HSC survey was limited by the twin effects of
[i] the finite apparent size of the source stars in its sky target, M31, being larger than the apparent Einstein radius,
yielding poorly focused light, and
[ii] transition from geometric to wave optics as the PBH Schwarschild radius becomes comparable to or smaller than the wavelength of optical light as the PBH mass is reduced.
Therefore, broadly speaking, to probe PBH masses less than $10^{-11}~M_\odot$ one must seek to microlens compact sources that emit in wavelengths smaller than optical.
Reference~\cite{baiorlofsky} recognized that accreting x-ray pulsars in the Magellanic Clouds provide just such a source, which moreover furnish a steady flux above which transient magnification of microlensing can be discerned, and are located far enough from Earth to provide appreciable optical depth of intervening PBHs. 
In that work, limits on atom-sized PBHs were derived from 10 days of observations at the Rossi X-ray Timing Explorer (RXTE) satellite on x-ray pulsars in the Magellanic Clouds, but these could only exclude the unphysical scenario of PBH densities higher than that of dark matter. 
Future projections of such other satellites as eXTP, AstroSat, Lynx, and Athena with 300 days of exposure were shown to be more promising.

In this work, we will show that the ongoing Neutron Star Interior Composition Explorer Mission (NICER)~\cite{NICERDesign2016}, designed primarily to measure the masses and radii of accreting x-ray pulsars, can constrain PBHs in the mass window as a physical fraction of the dark matter population with merely two months of live exposure.
This is comparable to the exposure already obtained by NICER on multiple sources.
Moreover, the future STROBE-X satellite, touted as the spiritual successor of NICER, would require even less exposure for this purpose thanks to its much larger effective area and wider coverage of x-ray frequencies.\footnote{A few months after our pre-print appeared, an announcement from NASA suggested that the STROBE-X proposal may not be funded~\cite{strobexnasanews}. Nonetheless in this work we will use ``STROBE-X" to denote a telescope with its capabilities and derive its sensitivities; it is always possible that this mission, or a similar one, may be revived.}
We estimate the reaches of these telescopes as a proof of concept that the observation of bright x-ray pulsars provide our best opportunity to discover dark matter in the PBH mass window of $10^{-16}-10^{-11}~M_\odot$.
While our analysis is not the most statistically rigorous due to poor understanding of systematics and correlations, our goal is to point out the feasibility of a microlensing search by these collaborations.
One main message of our paper is that {\em the science case is strong enough to fly a dedicated x-ray satellite for microlensing}.
Such a telescope should have a large effective area of about 1--10~m$^2$ and operate in the 0.1--1000 keV 
energy range, achievable traits as seen in the designs of the proposed instruments STROBE-X~\cite{STROBE-XScienceWorkingGroup:2019cyd}, LOFT-P~\cite{LOFTP2016} (which also may not materialize) and Daksha~\cite{dakshabhalerao2022,*dakshabhalerao2022science}. 
It should also continuously observe source candidates for a few weeks or months.
The costs of such a campaign are completely warranted given the monumental nature of discovering dark matter in an untouched region spanning five orders of magnitude.
In the rest of the paper, we will call this hypothetical future telescope ``X$\mu$", a name to connote x-ray microlensing.

This paper is laid out as follows.
In Section~\ref{sec:basics} we review the framework of gravitational microlensing by point-like lenses, with emphasis on x-ray pulsar sources and the effects of wave optics and finite source.
Here we discuss the identification of microlensing-induced magnification in the photon count rate data obtained at x-ray telescopes, and provide a diagnostic for distinguishing microlensing events from transient backgrounds, namely, the inspection of photon frequency data.
In Section~\ref{sec:limits} we derive event rates and summarize our prescription for setting microlensing limits in x-ray telescopes, using which we obtain limits on the population of PBHs with currently available NICER data, and future reaches of NICER, STROBE-X and X$\mu$. 
In Section~\ref{sec:discs} we provide discussion on the scope of our study.
In the appendix we provide some background material for key formulae used in the paper, and briefly survey alternative ideas in the literature for probing the PBH mass window.
We generally use natural units, $\hbar = c = 1$, but in some expressions we will restore $\hbar$ and $c$ for clarity.

%%%%%
\begin{figure*}[t]
    \centering
           \includegraphics[width=0.47\textwidth]{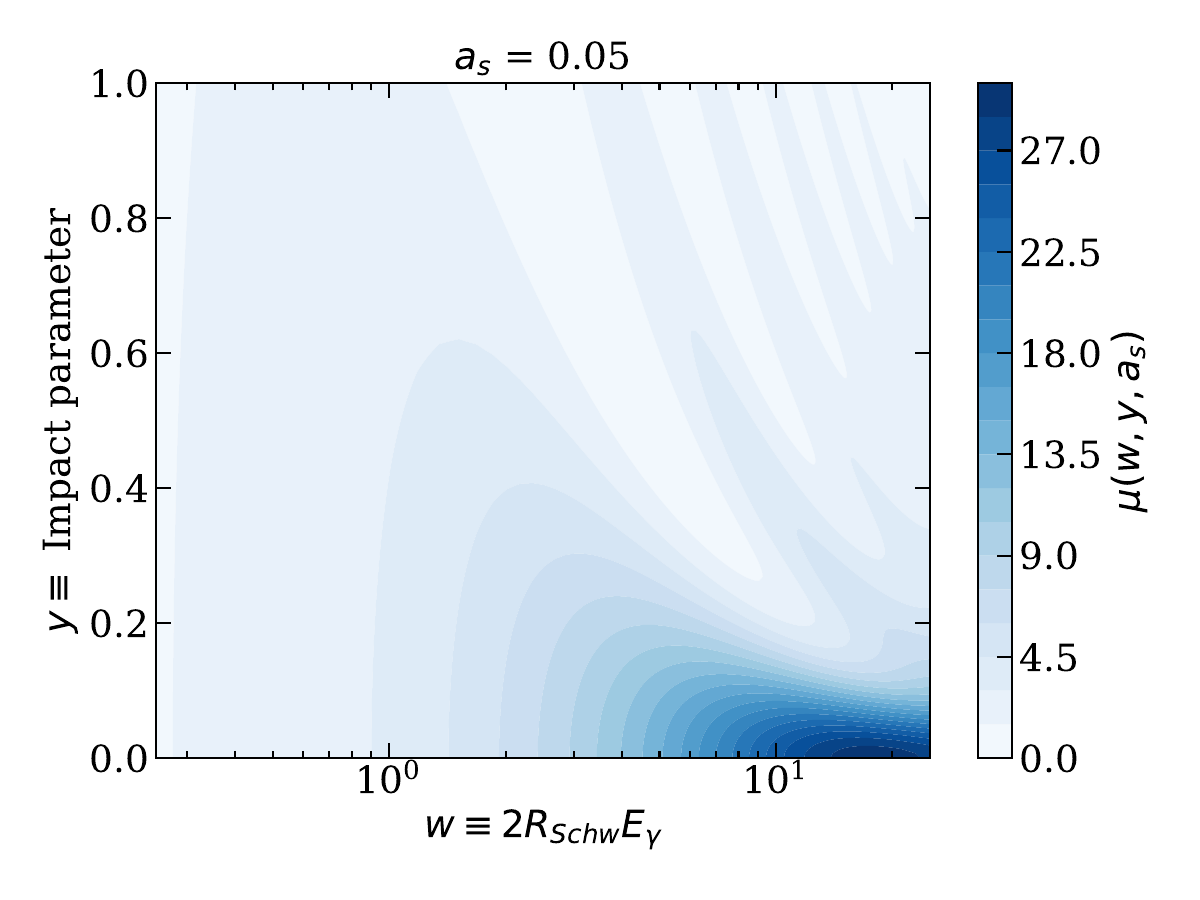}
           \includegraphics[width=0.47\textwidth]{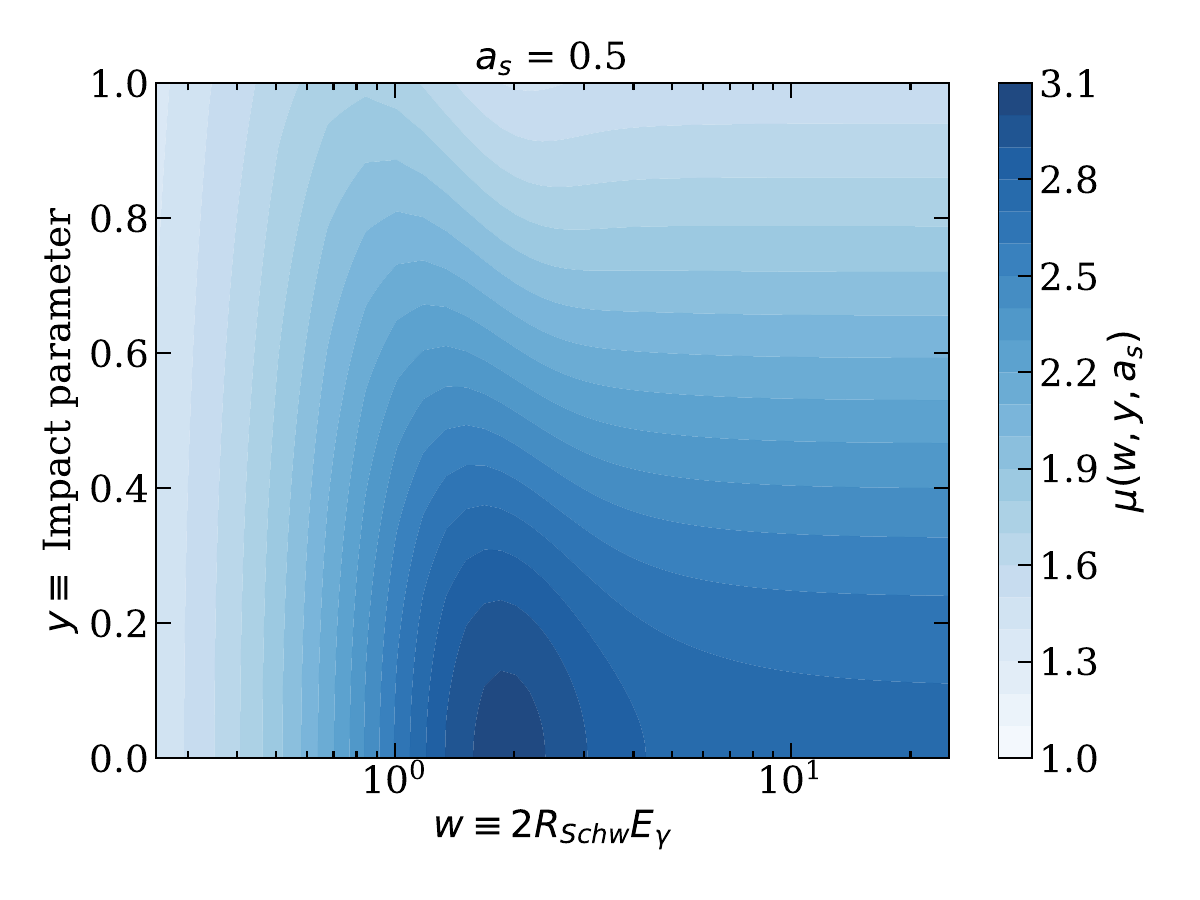}
    \caption{Contours of the magnification factor in Eq.~\eqref{eq:muwaveFSE} for point-like lenses in the plane of the wave/geometric optics demarcation parameter $w$ (Eq.~\eqref{eq:defn:w})  and the impact parameter $y$, for the source sizes $\aSource$ as indicated.
    See Sec.~\ref{sec:basics} for a discussion of the features here.
    }
    \label{fig:contoursmuwyx1}
\end{figure*}
%%%%%

%%%%%
\begin{figure*}[t]
    \centering
           \includegraphics[width=0.47\textwidth]{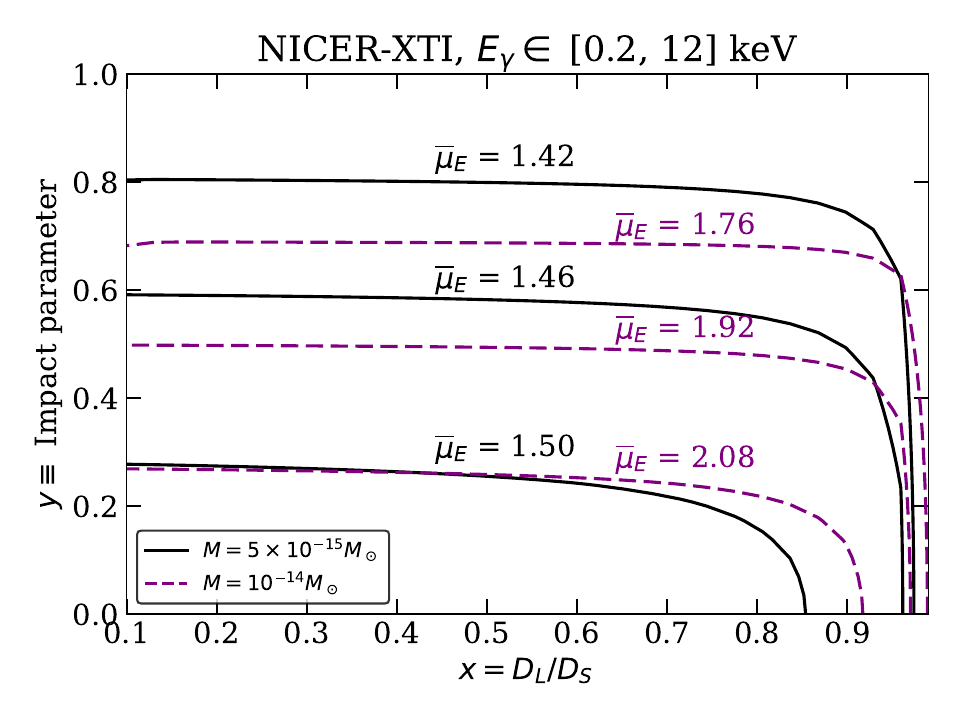}
           \includegraphics[width=0.47\textwidth]{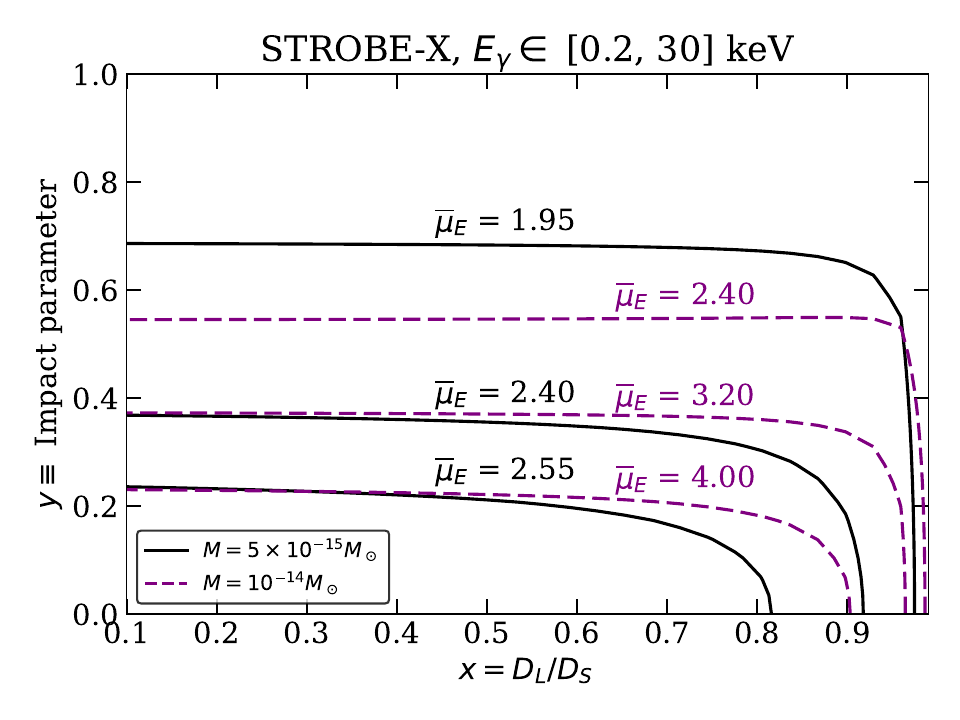}
    \caption{Contours of the source-specific energy-averaged magnification factor in Eq.~\eqref{eq:muenergyAv} for the x-ray pulsar SMC X-1 (source size: 20 km, distance from Earth: 64 kpc) at NICER and STROBE-X, with their detection energy ranges as indicated.
    See Sec.~\ref{sec:basics} for a discussion of the features here.
    }
    \label{fig:contoursmuwyx2}
\end{figure*}
%%%%

%%%%%%%%%%%%%%%%%%%%%%%%%
\section{X-ray Microlensing}
\label{sec:basics}
%%%%%%%%%%%%%%%%%%%%%%%%%

%%%%
\subsection{Basic set-up}
\label{subsec:waveFSE}
%%%%

The geometric set-up of microlensing is illustrated in, e.g.,  Refs.~\cite{NarayanBartelmann,CroonMcKeenRajECOlocation1,CroonMcKeenRajECOlocation2},
and microlensing of x-ray pulsars has been described in detail in Ref.~\cite{baiorlofsky}. 
Here we briefly review the essentials, consigning to Appendix~\ref{app:microlensingderivs} derivations of formulae.

Defining $x\equiv \DLens/\DSource$ as the ratio of the observer-lens distance $\DLens$ to observer-source distance $\DSource$,
the Einstein radius for a point-like lens of mass $M$ is given by
%%%
\bea
\label{eq:rE}
 \rE &=& \sqrt{\frac{4 G M x (1-x) \DSource}{c^2}} \\
\nn &=& 54~{\rm km} \bigg[ \bigg( \frac{M}{10^{-15} M_\odot} \bigg) \bigg(\frac{\DSource}{65~{\rm kpc}}\bigg) \bigg(\frac{x(1-x)}{0.25} \bigg) \bigg]^{1/2}~.
\eea
%%%%
This quantity is the closest distance that source photons get to the lens when it lies directly along the line of sight.
As line-of-sight distances of interest are generally much greater than  transverse distances in the problem, microlensing events may be visualized as projections on the lens-containing transverse plane, the ``lens plane". 
It then becomes useful to express distances in
units of $\rE$. 
The source radius in the lens plane is $\aSource(x) \equiv x \RSource/\rE$, and (in units of $\rE$) the distance from the lens center to the source center (i.e., the impact parameter) is $y$. 
As a function of these quantities, the magnification obtained at a telescope averaged over the energy range [$E_{\rm min}, E_{\rm max}$], weighted by the spectral energy distribution (SED) of the source $\mathcal{F}(E)$ and the energy-dependent effective area of the telescope $\mathcal{A}(E)$ is  
%%%
\beq
    \overline{\mu}_E (y, \aSource (x)) = \frac{\int_{E_{\rm min}}^{E_{\rm max}} dE \mathcal{A}(E) \mathcal{F}(E) \mu (w, y, \aSource(x))}{\int_{E_{\rm min}}^{E_{\rm max}} dE \mathcal{A}(E) \mathcal{F}(E)}~,
    \label{eq:muenergyAv}
\eeq
%%%
where 
%%%
\beq
\begin{split}
   & \mu(w,y,\aSource(x)) = \aSource^{-2} e^{-y^2/(2\aSource^2)} \frac{\pi w}{1 - e^{-\pi w}} \times \\
    & \int_0^{\infty} dz z e^{-z^2/(2\aSource^2)} I_0(yz/\aSource^2) \bigg|{}_1F_1\bigg(\frac{i w}{2}, 1; \frac{i w z^2}{2}\bigg)\bigg|^2 
    \label{eq:muwaveFSE}
\end{split}
\eeq
%%%
is a magnification factor that accounts for effects of both the finiteness of the source and wave optics, with ${}_1F_1$ the confluent hypergeometric function of the first kind. 
(We use the Python package {\tt mpmath}~\cite{mpmath} to evaluate ${}_1F_1$.)
The intensity of the source is here assumed to follow a Gaussian profile $\propto \exp(-|\vec{y} - \vec{z}|^2/2\aSource^2)$, and $I_0(z) = (1/2\pi) \int_0^{2\pi} e^{z cos\theta} d\theta $ is the zeroth-order modified Bessel function of the first kind~\cite{Nakamura:1999:waveoptix}.

The effect of wave optics is encapsulated by the parameter 
%%%%
\bea
\label{eq:defn:w}
w &&\equiv \frac{4 G M E_\gamma}{\hbar c^3} = \frac{2 \Rschw E_\gamma}{\hbar c}~\\
\nn \Rightarrow E_\gamma &&= 33.5~{\rm keV} \bigg(\frac{10^{-15} M_\odot}{M} \bigg)~w~,
\eea
%%%%
where $\Rschw = 2GM/c^2$ is the Schwarzschild radius of the PBH and $E_\gamma$ is the photon energy.
In the limit of $w \gg y^{-1}$ (where the impact parameter $y$ is typically $\Oc(1)$)
the extent of the lens well exceeds the photon wavelength, and Eq.~\eqref{eq:muwaveFSE} reduces to the expression obtained from geometric optics~\cite{Nakamura:1999:waveoptix,Matsunaga2006FiniteSourceWaveOptix,SugiyamaWaveptics:2019dgt,Montero-Camacho:2019jte} as seen by inspecting the behavior of ${}_1F_1$.
In the limit of $w \lsim y^{-1}$ wave optics effects become important, and the magnification tends to be suppressed for $w \ll y^{-1}$.
Thus Eq.~\eqref{eq:defn:w} also roughly marks the smallest PBH mass that can be probed by an instrument, corresponding to the largest energy it can detect, beyond which wave effects degrade the sensitivity.
Thus for \{NICER, STROBE-X/LOFT-P, RXTE\}, which reach energies of \{12, 30, 60\}~keV, the wave effects tend to suppress microlensing signals for $M < \{28, 11, 5.6\}\times 10^{-16}~M_\odot$.
Our proposed detector X$\mu$ would reach 1000 keV, hence PBH masses down to $3.4 \times 10^{-17}~M_\odot$. 
The exact value of $M$ down to which a given instrument can reach is determined by a few other factors, which will be discussed soon.

We depict the effects of finite source and wave optics in Figure~\ref{fig:contoursmuwyx1} with contours of the magnification factor in Eq.~\eqref{eq:muwaveFSE} in the space of $w$ and lens impact parameter $y$, for source sizes $\aSource$ of 0.05 and 0.5.
Scanning from left to right in both panels, we see that $\mu$ is suppressed for $w \ll 1$, deep in the wave optics regime, peaks at some intermediary $w$, and stabilizes at some $w \gg 1$ deep in the geometric optics regime.
Smaller impact parameters $y$ generally tend to produce larger magnifications, as expected.
Comparing across panels, the $\aSource = 0.5$ case typically produces smaller magnifications as expected: the more extended a source is on the lens plane, the less it is focused by the lens and hence the weaker the magnification~\cite{wittmao1994,CroonMcKeenRajECOlocation2}.
For $\aSource = 0.05$ large fluctuations in $\mu$ are visible in the region of transition between wave and geometric optics, $w = \Oc(1-10)$, a reflection of the rapid oscillation of the hypergeometric function in this region, best seen by taking appropriate limits of the $\aSource$-independent, $w$-dependent piece of the integrand in Eq.~\eqref{eq:muwaveFSE}~\cite{Matsunaga2006FiniteSourceWaveOptix}:
%%%%
\begin{widetext}
\beq
\frac{\pi w}{1 - e^{-\pi w}} \bigg|{}_1F_1\bigg(\frac{i w}{2}, 1; \frac{i w z^2}{2}\bigg)\bigg|^2 = 
\begin{cases}
1 + \frac{\pi w}{2} + \frac{w^2}{2}(\pi^2-3z^2) \ \ \ \ \ \ \ \ \ \ \ \ \ \ \ \ \ \ \ \ \ \ \ \ \ \ \ \ \ \ \ \ \ \ \ \ \ \ \ \ \ \ \ \ \ \  , w \ll 1~,~\\ \frac{1}{z\sqrt{4+z^2}}\bigg[2 + z^2 + 2 \sin\bigg[ w\bigg(\frac{1}{2} z \sqrt{4 + z^2} + \log\bigg|\frac{\sqrt{4+z^2}+z}{\sqrt{4+z^2}-z} \bigg| \bigg) \bigg] \bigg], \ w \gsim z^{-1}~.   
\end{cases}
\label{eq:1F1termapprox}
\eeq
\end{widetext}
%%%%

The white bands where $\mu \to 0$ depict regions where there is complete destructive interference.
For $\aSource = 0.5$, however, these oscillations are not visible since they are averaged out quickly via the tempering effect of the finite source, captured in Eq.~\eqref{eq:muwaveFSE} by the modified Bessel function term in the integrand; see the detailed illustrations in Ref.~\cite{Matsunaga2006FiniteSourceWaveOptix}.

In Figure~\ref{fig:contoursmuwyx2} we show contours of the energy-averaged detector- and source-specific magnification in Eq.~\eqref{eq:muenergyAv} in the plane of the impact parameter $y$ and fractional distance to the lens $x$.
The source here is taken to be the x-ray pulsar SMC X-1 in the Small Magellanic Cloud at a distance $d = 64$~kpc, with the source size, i.e., the radius of the emission region, taken to be $\RSource = 20$~km as done in Ref.~\cite{baiorlofsky}.
The SED of this pulsar is taken as~\cite{SEDSMCX1}
%%%%%
\bea
\nn 
&& \mathcal{F}(E_\gamma) = \\ 
\nn && \begin{cases}
(E_\gamma/{\rm keV})^{-0.93}\ \ \ \ \ \ \ \ \ \ \ \ \ \ \ \ \ \ \ \ \ \ \ \ \ \ ,~E_\gamma \leq 6~{\rm keV}, \\
(E_\gamma/{\rm keV})^{-0.93} e^{-(E_\gamma- 6~{\rm keV})/7.9~{\rm keV}},~E_{\gamma} > 6~{\rm keV}.
 \end{cases}\\
 \label{eq:SEDSMCX1}
\eea
%%%%%

The left and right panels depict respectively the currently operational X-ray Timing Instrument (XTI) on NICER and the forthcoming STROBE-X satellite, spanning energy ranges of [0.2, 12]~keV and [0.2, 30]~keV, whose effective areas $\mathcal{A}(E)$ are given in Ref.~\cite{LOFTP2016}.
We display two sets of contours corresponding to PBH masses $5 \times 10^{-15}~M_\odot$ and $10^{-14}~M_\odot$, having Schwarschild radii of ~(13.4 keV)$^{-1}$ $\equiv 1.5\times10^{-2}~$nm and (6.7 keV)$^{-1}$ $\equiv 2.9\times10^{-2}$~nm respectively. 

For a given $y$ the values of $\mubarE$ obtained at STROBE-X are seen to be generically larger than at NICER.
This is because STROBE-X can reach higher photon energies than NICER, so that it samples more of the geometric optics-dominated (as opposed to wave optics-dominated) region of microlensing, leading to overall stronger magnification.
In both panels the heavier PBH obtains higher magnification for comparable $(x, y)$ values, as once again it lies more in the geometric optics region, i.e., $w$ is larger for a given photon energy.
As expected, the magnification increases as $y$ is reduced.
An interesting effect is the rate of variation of $\mubarE$ across $y$, which is faster (slower) for the heavier (lighter) PBH. 
This is once again due to the wave vs geometric optics at play.
As seen from Fig.~\ref{fig:contoursmuwyx1}, the magnification varies slowly with $y$ in the wave optics ($w \ll y^{-1}$) region and more quickly in the geometric optics ($w \gg y^{-1}$) region.
The contours in Fig.~\ref{fig:contoursmuwyx2} rapidly go to $y = 0$ beyond some value of $x = x_{\rm max}$. 
This is the result of the finite source effect, which suppresses the magnification for lenses too close to the source.
If larger  $\mubarE$ values are desired, it comes with the price of smaller $x_{\rm max}$ in order to obtain a small source size on the lens plane.

%%%%%
\begin{figure*}[thb] \centering
    \includegraphics[width=0.5\textwidth]{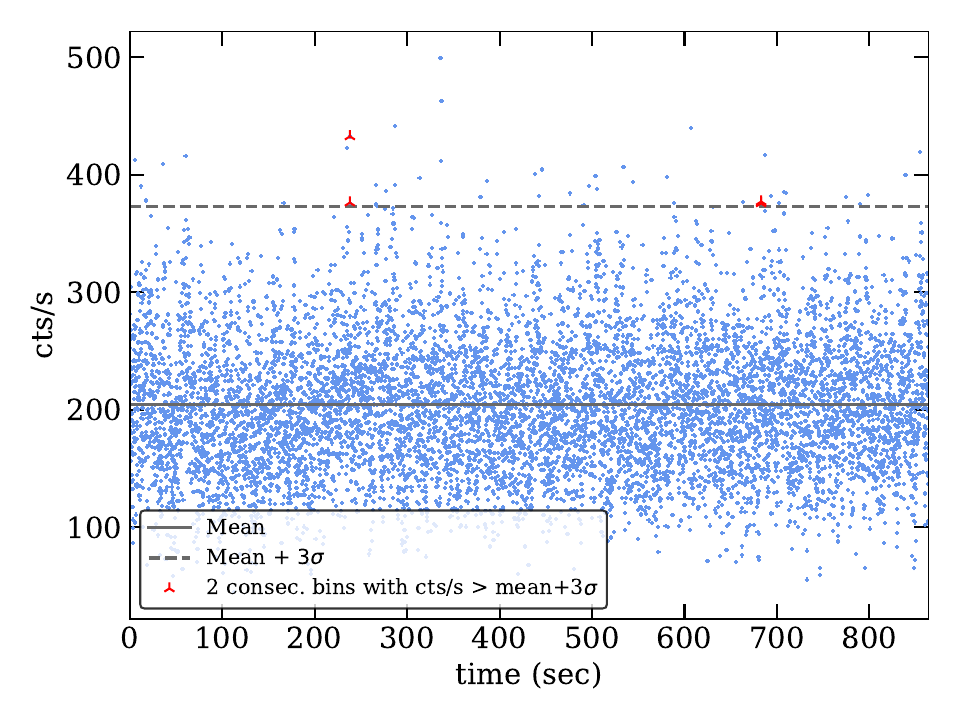} \\
          \includegraphics[width=0.8\textwidth]{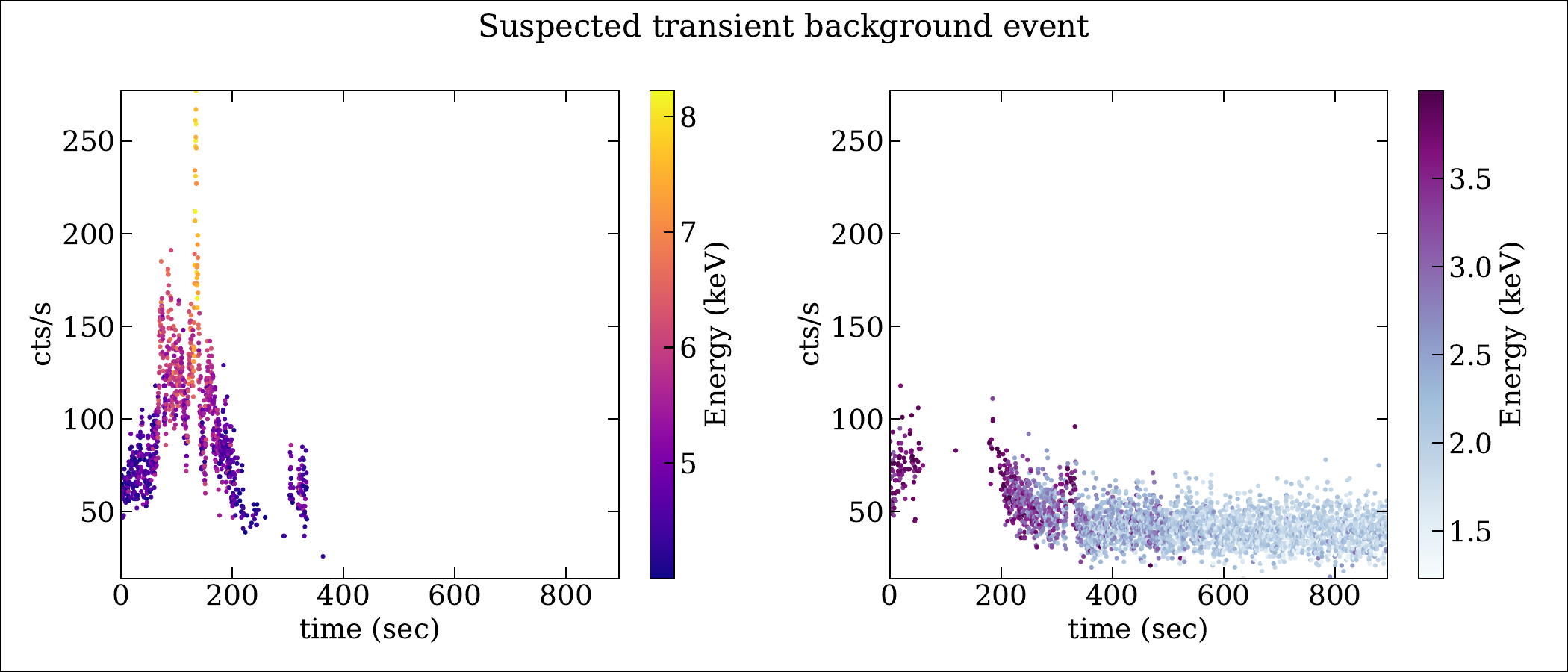}
    \caption{
     {\bf \em Top}: An 864-second segment (observation ID = 6535010301) of the x-ray pulsar SMC-X1 at NICER-XTI, binned by 0.1 seconds. With horizontal lines are shown the mean $B$ and the count rate 3 standard deviations $\sigma_B$ from it, and with red crosses are shown regions with 2 consecutive bins with count rate greater than $B + 3 \sigma_B$.
    {\bf \em Bottom}: Another segment (observation ID = 6535010801) showing a transient event.
    To confirm this as an occurrence unrelated to microlensing, such as an x-ray flare or interaction of atmospheric electrons with the detector, we inspect the event in two windows of energy (here with 0.2 second time-binning).
    Dot colors depict the average photon energy in a time bin, whose size we take as 0.2 seconds here. 
    Microlensing induced by PBHs in the geometric optics regime would be achromatic and thus appear the same in the two windows.
    See Sec.~\ref{subsec:spectraldiagnostic} for further details.}
    \label{fig:nicerlc}
\end{figure*}

%%%%%%

%%%%%%
\subsection{Identifying microlensing in x-ray pulsar data}
\label{subsec:spectraldiagnostic}
%%%%%%

Having reviewed microlensing in the x-ray regime, we now turn to how microlensing events may be identified in data.
We will make use of information on both count rates and energies of the photons detected at x-ray telescopes.

To uncover a PBH transit producing excess photon counts such as in microlensing, we follow the statistical prescription in Ref.~\cite{baiorlofsky}.
Assuming that time-binned photon counts are independent events and follow a Gaussian distribution, we demand that there are $N_{\rm consec}$ bins with photon counts that are at least $N_\sigma$ standard deviations ($\sigma_B$) greater than the mean counts per second $B$.
The values of $N_{\rm consec}$ and $N_\sigma$ must be chosen prudently so that the signal is statistically significant, i.e., the probability that such bins are produced by the statistical fluctuations must be sufficiently minuscule.
We will soon show that $N_{\rm consec} = 3$ and $N_\sigma = 3$ are optimal choices, which we will use for the rest of this work.

We must also ensure that a magnified signal (as defined below) has not simply arisen from a large number of statistical samples, i.e., from the look-elsewhere effect. 
The probability of obtaining 3 consecutive events above the $3 \sigma$ threshold should be smaller than 
%%%%
\beq
\frac{t_{\rm bin}}{t_{\rm exp}} = 1.9 \times 10^{-8} \bigg(\frac{t_{\rm bin}}{0.1~{\rm s}}\bigg) \bigg(\frac{60 \ {\rm days}}{t_{\rm exp}}\bigg)~, 
\eeq
%%%%
where $t_{\rm exp}$ is the exposure time. 
This probability $\mathbb{P}$ indeed satisfies the above condition for the 60-day exposures we consider, as it is $\mathbb{P}(N_\sigma) = (1 - \Phi(N_\sigma))^{N_{\rm consec}} = 2.5 \times 10^{-9}$, where $\Phi(x) = \frac{1}{2} \bigg(1 + {\rm erf} (x/\sqrt{2})\bigg)$ is the Gaussian cumulative distribution function (CDF)~\cite{baiorlofsky}. 
This is 7.5 times smaller than $t_{\rm bin}/t_{\rm exp}$. 

The minimum energy-averaged magnification required to detect microlensing events at a given telescope,
for count rates that are $n_\sigma \sigma_B$ away from the mean, is given as 
%%%%%
\beq
\muthresh = \frac{1 + N_\sigma \sigma_{B}/B}{1 + n_\sigma \sigma_{B}/B}~.
\label{eq:muthresh}
\eeq
%%%%%
As we will see in Sec.~\ref{sec:limits}, a conservative and reasonable choice of $n_\sigma$ is -1, which we will adopt. 
Variations in binning time $t_{\rm bin} \to \tilde t_{\rm bin}$ and bin counts $B \to \tilde B$ can be incorporated in Eq.~\eqref{eq:muthresh} by the replacement $\sigma_B/B \to (\sigma_B/B)/\sqrt{(\tilde B/ B)(\tilde t_{\rm bin}/t_{\rm bin})}$.

To justify our choices of $N_{\rm consec}$, $N_\sigma$ and $n_\sigma$ visually, we show in Figure~\ref{fig:nicerlc} an 864-second sample, with $t_{\rm bin}$ = 0.1 sec, of the NICER-XTI data on SMC-X1 extracted using {\tt HEASoft} v6.33.2~\cite{HEASoft} with {\tt nicerl2} and {\tt nicerl3-lc} pipelines. 
In this sample the mean $B = 204.26$ cts/s, which is indicated with a solid horizontal line, and the standard deviation $\sigma_B = 56.23$ cts/s; the 3$\sigma_B$ deviation from the mean is indicated with a dashed horizontal line.
We mark with red crosses bins that show 2 consecutive counts that are more than 3$\sigma_B$ from the mean.
We find no regions with 3 consecutive bins satisfying the criterion in any of the observation segments used in our analysis, and therefore use $N_{\rm consec} = 3$. 
Higher values of $N_{\rm consec}$ would diminish the probability $\mathbb{P}$ and thus the number of signal events (see Eq.~\eqref{eq:Nevents}); e.g., for $N_{\rm consec} = 4$, $\mathbb{P}$ is smaller than for $N_{\rm consec} = 3$ by a factor of at least 100 for any fixed value of $N_\sigma$.
Now for $N_{\rm consec} = 3$, we find that $\mathbb{P}(N_\sigma = 2) = 1.2 \times 10^{-5}$, $\mathbb{P}(N_\sigma = 3) = 2.5 \times 10^{-9}$, and $\mathbb{P}(N_\sigma = 4) = 3.2 \times 10^{-14}$.
Picking $N_\sigma = 4$ would again be too stringent a criterion, setting the value of $\mu_{\rm thresh}$ too high and thus diminishing the number of microlensing events (Eq.~\eqref{eq:Nevents}).
Picking $N_\sigma = 2$ would relax this criterion, but as seen from the top panel of Fig.~\ref{fig:nicerlc}, would also incur considerable background (unlensed) events.
Thus $N_\sigma = 3$ is a favorable middle ground that we pick.

%%%%%%
% For $N_{consec} = 2$, $\mathbb{P}(N_\sigma = 2) = 5.2 \times 10^{-4}$, $\mathbb{P}(N_\sigma = 3) = 1.8 \times 10^{-6}$, and $\mathbb{P}(N_\sigma = 4) = 1.0 \times 10^{-9}$; \\
% for $N_{consec} = 3$, $\mathbb{P}(N_\sigma = 2) = 1.2 \times 10^{-5}$, $\mathbb{P}(N_\sigma = 3) = 2.5 \times 10^{-9}$, and $\mathbb{P}(N_\sigma = 4) = 3.2 \times 10^{-14}$; \\
% for $N_{consec} = 4$, $\mathbb{P}(N_\sigma = 2) = 2.7 \times 10^{-7}$, $\mathbb{P}(N_\sigma = 3) = 3.3 \times 10^{-12}$, and $\mathbb{P}(N_\sigma = 4) = 1.0 \times 10^{-18}$; \\
%%%%%%

Note that $\sqrt{B t_{\rm bin}} = 4.52$ is smaller than $\sigma_B t_{\rm bin} = 5.62$, implying that the assumption of Gaussian statistics is not entirely correct and that there is some intrinsic variability in the count rate.
This may also be inferred from the fact that there are 3 red crosses visible (and one hidden behind the third cross) in the top panel of Fig.~\ref{fig:nicerlc}, whereas the probability that 2 consecutive events are seen $3 \sigma$ above the mean is only about 1\%.
Moreover, we find that the probability distribution function of counts is skewed a little to the right of a Gaussian.
Further, the fact that the red crosses are seen to cluster across bins (as also seen in Ref.~\cite{baiorlofsky}) suggests that the events are correlated across bins, for instance due to micro-flares.
These imply that the limits we obtain are on the optimistic side, and a more careful likelihood procedure must be followed by microlensing searches, including the empirical count probability distribution function and known correlations across events.
With these caveats, the threshold magnification in Eq.~\eqref{eq:muthresh} is then $\muthresh = 2.5$.

The count rate, and hence the threshold magnification in Eq.~\eqref{eq:muthresh}, is determined by the source SED as well as the effective area of the detector.
For our proposed telescope X$\mu$, we assume an energy-dependent effective area that is a combination of three parts:
[i] for $E \in [0.2, 50]$~keV 
we adopt the $\mathcal{A}(E)$ of the Large Area Detector (LAD), to be used on STROBE-X~\cite{STROBE-XScienceWorkingGroup:2019cyd} and eXTP~\cite{eXTPferoci2018large};
[ii] for $E \in [50, 260]$~keV we take the $\mathcal{A}(E)$ profile of the cadmium zinc telluride ``medium energy" detector proposed to be used in Daksha~\cite{dakshabhalerao2022,*dakshabhalerao2022science}, and scale the normalization up by a factor of 2,
[iii] for $E \in [260, 980]$~keV we take the $\mathcal{A}(E)$ profile of the NaI ``high energy" detector of Daksha and again rescale by a factor of 2.
The energy-averaged detector area weighted by the source SED, $\overline{\mathcal{A}} = \int_{E_{min}}^{E_{max}} dE \mathcal{F}(E)\mathcal{A}(E) / \int_{E_{min}}^{E_{max}} dE \mathcal{F}(E)$, is 30.9 and 46.8 times bigger for STROBE-X and X$\mu$ respectively as compared to NICER.
Rescaling $\muthresh$ suitably by $\sqrt{\overline{\mathcal{A}}}$ we obtain $\muthresh = 1.21$ for STROBE-X and $\muthresh = 1.17$ for X$\mu$, which we use for obtaining projections in Sec.~\ref{sec:limits}.

In selecting observation samples for analysis, care must be taken to remove bins that are contaminated by x-ray flares and bursts, as well as atmospheric backgrounds such as trapped electrons (TREL), precipitating electrons (PREL), and low-energy electrons (LEEL)~\cite{SCORPEON}.
This may be done manually by visual inspection for distinct rise-and-falls in photon counts, but to ascertain that we are not discarding microlensing events, we also perform the following procedure.
Microlensing in the geometric optics regime (i.e., for PBHs larger than the photonic wavelength) is achromatic: it produces source frequency-independent magnification.
We exploit this fact to look for similarities in the light curve in two different energy windows for suspected flares or PREL events.
The bottom panel of Figure~\ref{fig:nicerlc} shows an example of a suspected x-ray flare in the SMC-X1 NICER data, with either panel displaying photon counts with energies $>$ and $\leq$ 4 keV.
(To obtain this plot we use information on individual photon energies and detection timestamps, and then show the count rate in 0.2 sec bins.)
If this had been a microlensing event, these panels would have looked alike, but they look extremely dissimilar.
That the higher-energy counts alone occur as an excess confirms that this event is indeed an x-ray flare or burst.
It is still possible that when excising these flares, a PBH microlensing event may get discarded, but it is unlikely that such an event coincides in time with a flare.
As such, our limits become conservative in the process of removing bins from the final analysis. 
While we have removed obvious flares by visual inspection, a microlensing search must use a more careful algorithm to remove smaller flares that may be close to detection threshold.
We add here that, due to the uncertain nature of these backgrounds, claiming a discovery would be much more challenging than setting constraints in real experiments.

A diagnostic like this serves not only to distinguish microlensing events from background transients, but to also roughly inform us of the mass of the PBH should a true microlensing event be observed.
If the count rate vs time of a suspected signal event corresponds to the characteristic shape of the light curve expected from microlensing, and is seen to be identical -- or more practically, statistically consistent -- in two or more energy windows, then we may be sure that the transiting PBH is massive enough to be in the geometric optics regime, where microlensing is indeed achromatic. 
The geometric vs wave optics nature of microlensing may be further verified by extracting the lens mass, at least broadly, by:

[i] comparing the event time scale with the lens transit time, roughly 2$\rE$ divided by the transverse speed of dark matter.
Equation~\eqref{eq:rE} gives $\rE$ in terms of $M$ and $x$, if the dark matter speed is taken as its typical value (about 200 km/s);

[ii] fitting the light curve to obtain the magnification, which contains further information about $M$ and $x$ via Eq.~\eqref{eq:muenergyAv}.

The range of $x$ can be further narrowed to some $\Oc(0.1)$ value by 
considering its probability distribution, which is the differential optical depth $\pi \rE^2(x) n_{\rm PBH}(x) dx$, where $n_{\rm PBH}$ is the PBH number density along the line of sight to the source.

If the PBH mass is in the wave optics regime, confirmation of microlensing would be unfortunately more complicated.
This is because in this regime the microlensing magnification {\em is} frequency-dependent, as discussed in Sec.~\ref{subsec:waveFSE} and in, e.g., Refs.~\cite{Nakamura:1999:waveoptix,Matsunaga2006FiniteSourceWaveOptix,SugiyamaWaveptics:2019dgt}.
An additional complication is that magnifications are suppressed for impact parameters $y > w^{-1}$, as seen in Sec.~\ref{subsec:waveFSE}.
Also, the signal would be initially and finally achromatic during the transit but chromatic when the lens is nearest to the source on the lens plane. 
This may be seen from Eq.~\eqref{eq:1F1termapprox}, where the sine term $\to 0$ for fast oscillations.
In this regime the detector energy resolution becomes important: if it is smaller than the change in $w$ over which the magnification varies appreciably for a suspected lens mass, chromatic microlensing may be measured.
Perhaps astrometric microlensing may lift the degeneracies discussed here, as demonstrated for microlensing in optical frequencies~\cite{astrometricmass2020,*astrometricmassOGLE:2022gdj}, but this could be challenging for x-ray microlensing due to uncertainties in pulsar emission and localization.
We leave the question of diagnosing microlensing signatures in the wave optics regime to future investigation.

%%%%%%%%%%%%%%%%%%%%%%%%%%
\begin{table*}[t]
\begin{center}
\begin{tabular}{ l c c c c } 
\hline
x-ray pulsar  & net exposure (days) & $\DSource$ (kpc) & ($\ell$, $b$) & $\sigma_B/B$  \\
\hline
\hline
SMC X-1 & 1.74 & 64~\cite{SMCX1Dist2005}     & $(300.41^{\circ}, -43.56^{\circ})$ & 0.28  \\
\hline
Cyg X-2 & 5.47 & 11~\cite{Ludlam:2022yjp}  & $(87.33^{\circ}, -11.32^{\circ})$ & 0.02 \\
\hline
Vela X-1  & 4.46 & ~2~\cite{vallenari2023gaia} & $(263.06^{\circ}, 3.93^{\circ})$ & 0.25 \\
\hline
Crab pulsar & 4.76  & ~2~\cite{CrabLin2023}  & $(184.56^{\circ}, -5.78^{\circ})$ & 0.01  \\
\hline
\end{tabular}
\end{center}
\caption{X-ray pulsars in the NICER dataset~\cite{NICERDataTables} with significant count rates and small intrinsic variability $\sigma_B/B$, and the net telescope exposure on them. Also given are their distances from Earth and location in Galactic co-ordinates.
Due to its distance SMC X-1 provides the greatest microlensing optical depth, thereby dominating the total event rate.
See Sec.~\ref{sec:limits} for further details.}
\label{tab:NICER-psr}
\end{table*}
%%%%%%%%%%%%%%%%%%%%%%%%%%

%%%%%%
\section{Event rates and telescope sensitivities}
\label{sec:limits}
%%%%%%

Assuming that the PBH mass spectrum is monochromatic (i.e., that all PBHs have a single mass), and that their velocities follow a Maxwell-Boltzmann distribution, the differential rate of microlensing with respect to event timescale $\tE$ per source pulsar is given by~\cite{griest1991}
%%%%
\bea
   \nn \frac{d\Gamma}{d \tE} =&& f_{\rm PBH}  \frac{2 \DSource}{M v^2_0}  \int_0^1 dx \rhodm (x) \vE^4(x) e^{-\vE^2/v_0^2} \times \\
   && \int_0^{y_{\rm T}(x)} \frac{dy}{\sqrt{y_{\rm T}^2 - y^2}} \times \mathbb{P}(n_\sigma)~,
   \label{eq:μeventrate}
\eea
%%%%
where 
$\vE \equiv 2 \rE \sqrt{y_{\rm T}^2 - y^2}/\tE$, with $y_{\rm T}(x)$ the maximum impact parameter within which the magnification $\geq \muthresh$ as illustrated in Fig.~\ref{fig:contoursmuwyx2},
$v_0$ is the halo circular velocity $\simeq 240$~km/s, 
$f_{\rm PBH}$ is the mass fraction of PBHs making up the dark matter density $\rhodm$, 
and $\mathbb{P}(n_\sigma)$ is the fraction of bins in the dataset with count rate higher than $B + n_\sigma \sigma_B$ for 3 consecutive bins.
We find that $\mathbb{P}(-1) = 0.596$ for a Gaussian CDF with $n_\sigma = -1$.
(For $n_\sigma = -2$ we have $\mathbb{P}(-2)  = 0.93$, which would be an aggressive choice, and for $n_\sigma = 0$ our event acceptance would be low with $\mathbb{P}(0) = 0.13$. 
One could also optimize the choice of $n_\sigma$ by marginalizing over the intrinsic fluctuations in the source's photon emission, but this is beyond our scope of determining order-of-magnitude sensitivities.)
Note that in general the first integral in Eq.~\eqref{eq:μeventrate} evaluates to non-zero values only up to some $x = x_{\rm max}$ since, as seen from Fig.~\ref{fig:contoursmuwyx2}, $y_{\rm T}(x) \to 0$ above $x_{\rm max}$, thereby making the second integral vanish.

We show results for two different spatial distributions of $\rhodm$, ignoring the line-of-sight DM density in satellite galaxies (where some of our pulsars are situated) as it contributes only about 10\% to the event rate. 
First we take the Einasto profile,
%%%%
\bea
\label{eq:Einasto}
\rho_{\rm Ein} (r) = \frac{M_0}{4\pi r_s^3} e^{-(r/r_s)^\alpha},\\
\nn r = \sqrt{r_\odot^2 + x^2 \DSource^2 - 2 r_\odot x \DSource \cos\ell \cos b}~,
\eea
%%%%
where
$r_\odot = 8.33$~kpc  is the distance of the Sun from the Galactic Center, and ($\ell$, $b$) are the Galactic co-ordinates of the source.
Recent studies~\cite{gaianecibvcirc2023} using photometry data from Gaia DR3, 2MASS and WISE combined with SDSS-APOGEE DR17 spectra 
show that Milky Way circular velocities for galactic radii $\leq$ 30 kpc are best fit by the above Einasto profile with a normalization mass $M_0 = 6.2\times 10^{10}~M_\odot$, 
scale radius $r_s = 3.86$~kpc, and
slope parameter $\alpha = 0.91$, which we adopt.
As a cored Einasto profile is found to better fit the circular velocities estimated in Ref.~\cite{gaianecibvcirc2023} than, say, a cuspy Navarro-Frenk-White (NFW) profile, the virial mass of the Milky Way turns out to be smaller than previous estimates, which is consistent with the findings of Refs.~\cite{JiaoKepleriandecline:2023aci,gaianecibvesc2024} that use Gaia DR3 data at smaller galactic radii. 
The upshot is that the use of Eq.~\eqref{eq:Einasto} makes dark matter densities, hence the microlensing optical depth, generally smaller than in earlier works such as Refs.~\cite{Niikura2019Subaru,CroonMcKeenRajECOlocation1,CroonMcKeenRajECOlocation2}.

Nevertheless, we also derive optimistic projections by assuming the NFW profile, 
%%%%%
\beq
\rho_{\rm NFW} (r) =  \frac{\rho_0}{(r/r_s)(1+r/r_s)^{2}}~,
\label{eq:NFW}
\eeq
%%%%%
to make direct comparisons with earlier literature and which may better describe the Galactic halo in future and/or complementary data.
To fix the unknown parameters in Eq.~\eqref{eq:NFW}, 
we 
[i] fix $\rho_\odot = 0.44$~GeV/cm$^3$ as the dark matter density at the solar position as obtained from the best-fit of Eq.~\eqref{eq:Einasto}; this ensures that dark matter densities at least near the Earth are comparable in the two profiles we consider,
[ii] take the mass of dark matter within $r$ = 60 kpc 
as $4.7 \times 10^{11}~M_\odot$ as estimated by the SDSS survey in Ref.~\cite{SDSS:2008nmx}.
These give $\rho_0 = 0.95$~GeV/cm$^3$ and $r_s = 11.46$~kpc.
Other choices that fix NFW parameters, as well as the best-fit generalized-NFW profile in Ref.~\cite{gaianecibvcirc2023}, yield constraints that are quite similar to those obtained with the best-fit Einasto profile.

The total number of microlensing events expected in a telescope, summing over pulsar targets labelled by $i$ with individual observation times $T^i_{\rm obs}$, is now
%%%%
\beq
N_{\rm ev} = \sum_i T_{\rm obs}^i \int_{t_{{\rm min},i}}^{t_{{\rm max},i}} d\tE \frac{d\Gamma_i}{d \tE}~,
\label{eq:Nevents}
\eeq
%%%%
where $t_{{\rm min},i}$~is the binning time (= 0.1 sec here) and $t_{{\rm max},i}$ is the net exposure for the source $i$.
These integration limits are chosen as the shortest and longest timescales over which a microlensing event can be observed at a telescope.
Thus constraints on the unknowns $f_{\rm PBH}$ and $M$ may be derived by comparing $N_{\rm ev}$ with the observed number of events under the no-PBH null hypothesis.
For example, if zero events are observed, using Poisson statistics the 90\% C.L. limit can be obtained by setting $N_{\rm ev} = 2.3$.
In practice, we evaluate the integral in Eq~\eqref{eq:Nevents} using the analytic form identified in the appendix of Ref.~\cite{CroonMcKeenRajECOlocation1}.

%%%%%
\subsection{Prescription for obtaining microlensing limits}
%%%%%
Before we discuss our main results, let us first summarize the series of steps for obtaining the constraints as described in the text up to this point.

(1) Select x-ray pulsar sources with high count rates, low intrinsic variability, and possibly large distances to maximize the PBH microlensing optical depth.

(2) Obtain data on count rate vs time and photon energies on the x-ray pulsar sources, combining segments such as those in Fig.~\ref{fig:nicerlc}. 
Use appropriate data-processing pipelines.

(3) Inspect the dataset for transient excesses. 
To verify them as non-microlensing events, check the count rate in two or more different photon energy windows such as in the bottom panel of Fig.~\ref{fig:nicerlc}.

(4) If the count rates in different energy windows are dissimilar, discard the corresponding time bins. 
If they are similar, investigate further if the shape of the light curve corresponds to a microlensing signature.
Attempt to break degeneracies in microlensing free parameters as discussed at the end of Sec.~\ref{subsec:spectraldiagnostic}.

(5) From the subset of data with chromatic excesses removed, obtain the mean and standard deviation of the count rate.
Optimize choices of $N_{\rm consec}$, $N_\sigma$ and $n_\sigma$.
 Then estimate the threshold magnification $\muthresh$ from Eq.~\eqref{eq:muthresh}.

(6) Use this $\muthresh$, the effective area $\mathcal{A}(E)$ and the source SED $\mathcal{F}(E)$ to obtain $y_{\rm T}$ vs $x$ from Eq.~\eqref{eq:muenergyAv} as in Fig.~\ref{fig:contoursmuwyx2}.

(7) Finally, use the microlensing event rate in Eq.~\eqref{eq:μeventrate} to estimate the total number of events expected (Eq.~\eqref{eq:Nevents}).
Use appropriate statistics to report limits at the desired confidence level.

%%%%%
\subsection{Results}
\label{subsec:results}
%%%%%
In Fig.~\ref{fig:fvM} we display with dashed (dotted) curves the NFW (Einasto) profile 90\% C.L. limits obtained with current data at NICER, with effective area $\Oc (10^{-2}-10^{-1})$~m$^2$, on the pulsars listed in Table~\ref{tab:NICER-psr}, after removing transient excesses caused by flares, electron backgrounds, etc.
These pulsars were selected for the appreciable photon count rates they yielded at NICER, bright as they are. 
Despite a somewhat lower exposure compared to Cyg X-2, Vela X-1 and the Crab pulsar, the SMC X-1 pulsar dominates the net microlensing event count in Eq.~\eqref{eq:Nevents} due to its large distance from Earth that results in a relatively high microlensing optical depth.
From SMC X-1 count rate data such as the sample in Fig.~\ref{fig:nicerlc}, using Eq.~\eqref{eq:muthresh} we get the threshold energy-averaged magnification $\muthresh = 2.5$.
We see that the data accumulated so far at NICER could only set $f_{\rm PBH} < 20$ in the mass range of about $10^{-14}-10^{-13}~M_\odot$. 
If the exposure on SMC X-1 were increased to 60 days, we see that NICER can reach $f_{\rm PBH} = 0.7$ for $M = 2 \times 10^{-14}~M_\odot$ and can generally probe the mass range $(1-5) \times 10^{-14}~M_\odot$ for $f_{\rm PBH}\leq 1$.
Such an exposure is justified by the enormous significance of the physics case here.
We also believe it is feasible as this exposure is comparable to that obtained at NICER for some sources, e.g., a week~\cite{largeexposure:Deneva:2019jal}, and to that of other sources, e.g., PSR B1937+21, as seen in the NICER data archive~\cite{NICERDataTables}.

We also show the 30-day reaches of future x-ray telescopes with effective areas about an order of magnitude greater than that of NICER.
For STROBE-X we use SMC X-1 as the source.
For X$\mu$ we assume a hard x-ray pulsar source that is at $\DSource = 64$~kpc, with intrinsic variability $\sigma_B/B$ that of SMC X-1 (as in Table~\ref{tab:NICER-psr}), and an energy spectrum that is the same as the Crab pulsar~\cite{SEDMeVPulsarsharding2017mev}, which provides a flux orders of magnitude higher than that of SMC X-1 for $10-1000$~keV energies.
By placing this source far away our count rate is $1/\DSource^2$-suppressed, but we gain in microlensing optical depth.
As mentioned below Eq.~\eqref{eq:SEDSMCX1}, these telescopes can also probe photon energies higher than at NICER.
We see in Fig.~\ref{fig:fvM} that X$\mu$ could probe much smaller PBH masses than NICER, down to $2.5 \times 10^{-17} M_\odot$, while STROBE-X can reach down to about $10^{-15}~M_\odot$.
This is because these telescopes, by virtue of detecting smaller x-ray wavelengths than NICER, can overcome the wave optics effect and observe microlensing by smaller/lighter PBHs.
Note that requiring an excess over 3 consecutive 0.1 sec bins (as below Eq.~\eqref{eq:muthresh}) implies requiring a transit time across $\rE$ of 0.3 sec, which is expected on average for $M > 10^{-15}~M_\odot$, whereas the X$\mu$ limits go to smaller lens masses.
This is possible by shrinking $t_{\rm bin}$ suitably which does not unduly affect the counts per bin due to the large collecting power of the telescope, nor the threshold magnification as per the discussion under Eq.~\eqref{eq:muthresh}.
As such, the microlensing event count may slightly increase if $t_{\rm min}$ in Eq.~\eqref{eq:Nevents} is set to the new $t_{\rm bin}$, but our X$\mu$ estimate uses $t_{\rm min} = 0.1$~sec and must be taken as a conservative projection.

STROBE-X and X$\mu$ also reach smaller $f_{\rm PBH}$ of $\Oc (0.1)$ than NICER with less run-time.
This is due to their larger effective areas that result in smaller threshold magnification, in turn giving greater event rates as per Eq.~\eqref{eq:μeventrate}. 
We see that the limits on $f_{\rm PBH}$ appear to be proportional to $\sqrt{M}$ for large PBH masses.
This may be broadly understood as follows.
The PBH velocities are roughly constant, so that the only $M$-dependence in Eq.~\eqref{eq:μeventrate} is in the term outside the integrals.
Since $\tE \propto \sqrt{M}$, Eq.~\eqref{eq:Nevents} implies that $f_{\rm PBH} \propto \sqrt{M}$ for a fixed $N_{\rm ev}$.
Our results may be compared with that of Ref.~\cite{baiorlofsky}, which shows that with 300 days of exposure the future eXTP satellite would reach down to $f_{\rm PBH} \simeq 0.1$, and Athena, Lynx and AstroSat to $f_{\rm PBH} \simeq 0.3$.
Shorter exposures such as we have used will weaken the sensitivities of Ref.~\cite{baiorlofsky} proportionally.
While Ref.~\cite{baiorlofsky} had set limits using 10 days of RXTE data (as shown in Fig.~\ref{fig:fvM}), NICER obtains stronger limits with only 1.74 days as its effective area profile covers energies at which the SED of SMC-X1 peaks.
As mentioned in Ref.~\cite{baiorlofsky} one may also consider data from XMM-Newton, Chandra and other x-ray telescopes but their exposures and effective areas are generally smaller than at NICER and RXTE, so we do not consider them in our analysis.
The imager at XMM-Newton, EPIC, has a better spatial resolution of 6 arcsec compared to NICER's 5 arcmin, implying that backgrounds from outside the source may be potentially mitigated, however XMM-Newton has not only a smaller effective area but also a much smaller timing resolution of $\mathcal{O}(10^{-2}-1)$~ms than the $\mathcal{O}$(ns) resolution of NICER.
Moreover, NICER is an instrument dedicated mainly to the study of x-ray pulsars.

%%%%%%%%%%%%%%%%%%%%%
\section{Discussion and scope}
\label{sec:discs}
%%%%%%%%%%%%%%%%%%%%%

In this work we have investigated the sensitivity of x-ray telescopes studying x-ray pulsars to microlensing by sub-atomic size primordial black holes.
Our results are summarized in Figure~\ref{fig:fvM}.
We have also described a spectral diagnostic to confirm a microlensing signal in the geometric optics regime, which can further be confirmed by measuring the timescale and magnification of events.
We re-emphasize that, while searches at the ongoing NICER and planned STROBE-X telescopes would be definitely exciting, commissioning a new, large, broadband x-ray satellite dedicated to gravitational microlensing surveys would be even more worthwhile in light of the wealth of fundamental physics to be mined.
More realistically, such a telescope would need to be proposed by the astrophysics community with other interesting science goals included.
Thus we envision X$\mu$ to be primarily a microlensing instrument, but also performing other important measurements that x-ray telescopes are capable of, for instance, precise measurements of the neutron star mass-radius relationship, black hole mass, spin and accretion flows, x-ray counterparts of multi-messenger transients, and so on~\cite{NICERDesign2016,STROBE-XScienceWorkingGroup:2019cyd}.
Further, due to the possibly large costs involved in the construction and launch of a broadband observatory, one could also consider a small new telescope targeting only the highest x-ray frequencies.
This could be realized by using, e.g., a rescaled version of the NaI ``high energy" component of the proposed Daksha mission as mentioned in Sec.~\ref{subsec:spectraldiagnostic}.
This would probe the lowest PBH masses that are out of the reach of NICER and Strobe-X, and may even uncover mechanisms of black hole decay beyond Hawking evaporation (that we discuss shortly), enhancing the case for testing fundamental physics.
Moreover, as described in Appendix~\ref{app:otherwindowfasteners}, there may be other experiments in the near to middle future that probe the PBH mass window using other physical principles -- implying that some version of our proposed X$\mu$ satellite performing x-ray microlensing would be a valuable complementary and cross-verifying probe.

X-ray microlensing in these telescopes would help to uncover not only point-like lenses such as PBHs,
but also, as mentioned in the Introduction, dark matter in structures with extent comparable to the (point-like) Einstein radius, which would produce non-trivial magnification curves.
It may even be possible to distinguish between the possibilities, such as microhalos of various density distributions, boson stars, etc., via machine learning techniques~\cite{CrispimRomao:2024nbr}.
Whatever be the type of lens, mass functions other than the simple monochromatic one assumed here may be used to set limits, such as done in Ref.~\cite{LehmannMassFunc:2018ejc,*GortonMassFunc:2024cdm}.

Our work is the first to point out that the PBH evaporation limit at around $10^{-16}~M_\odot$, the lower end of the PBH mass window~\cite{Saha:2024ies}, may be probed with a minimal microlensing setup involving hard x-ray pulsars; see Appendix~\ref{app:otherwindowfasteners} for more sophisticated setups.
But the PBH mass window itself may be wider than previously supposed.
If black hole evaporation, which has never been observed, proceeds at a rate slower than Hawking's prediction as may happen if the PBHs are magnetic~\cite{mPBH:Maldacena:2020skw,*mPBH:BanerjeeThalapillil:2024sao} or via the memory burden effect~\cite{Dvali:2018xpy,Dvali:2020wft,Dvali:2024hsb}, then the extant limits on PBH masses from evaporation (at $M \gsim 10^{-16}~M_\odot$~) may be greatly weakened~\cite{Alexandre:2024nuo}.
It is even possible that PBHs as light as $10^{-28}-10^{-23}~M_\odot$ are allowed to comprise all the dark matter~\cite{Thoss:2024hsr,Dvali:2024hsb}.
To catch very light PBHs in microlensing would require sources emitting photons with wavelengths even smaller than x-rays.
Perhaps gamma-ray pulsars shining copiously to give sufficient statistics at such detectors as Fermi-LAT~\cite{Fermi-LAT:2023zzt} and its larger successors may be valuable.
Even in the absence of evaporation-slowing effects, 
PBH microlensing campaigns involving hard x-ray sources and gamma-ray pulsars would be worthwhile: they automatically constrain non-PBH compact dark matter structures that are smaller than the relevant Einstein radii.
In this light our X$\mu$ reach in Fig.~\ref{fig:fvM} at low $M$ must be read by ignoring the evaporation constraints.

All told, the time has come to coax dark matter out of one of its famed hideouts.

%%%%%%%
\section*{Acknowledgments}
%%%%%%%

We are indebted to Paul Ray for key insights on the operation of NICER and 
Abhisek Tamang for help with {\tt HeaSOFT}.
We further thank 
Varun Bhalerao,
Priyanka Gawade,
Ranjan Laha,
Surhud More,
Suvodip Mukherjee,
Avinash Kumar Paladi,
Akash Kumar Saha,
Vibhor Kumar Singh,
Abhishek Tiwari,
Ujjwal Kumar Upadhyay,
Himanshu Verma,
and
Anna Watts
for helpful discussion.
We also thank the anonymous referees for their careful assessment and inputs.

\appendix

%%%%%%%%%%%
\section{Background material on microlensing}
\label{app:microlensingderivs}
%%%%%%%%%%%

In this appendix we provide derivations for some formulae in Sec.~\ref{sec:basics}.
For detailed treatments of microlensing we refer the reader to Refs.~\cite{NarayanBartelmann,Matsunaga2006FiniteSourceWaveOptix}.

We start with some generic notions.
Consider the background metric $g_{\mu\nu}$ with a gravitational potential $U$ due to a point-like lens, given by
%%%%
\beq
    ds^2 \equiv g_{\mu\nu} dx^\mu dx^\nu = -(1 + 2U) dt^2 + (1 - 2U)d\vec{r}^2~.  
\eeq
%%%
As shown in Ref.~\cite{Takahashi2003GWLensingWaveFX}, for $U \ll 1$ we can treat the propagating degree of freedom of an electromagnetic wave as a massless scalar field $\phi$ obeying
%%%%
\begin{equation}
    \partial_\mu(\sqrt{-g}\hspace{1mm}g^{\mu \nu}\partial_\nu \phi) = 0~.
\end{equation}
%%%%
Setting $\phi(\vec{r},t) = \Phi(\vec{r})e^{-i \omega t}$, we get
%%%%%
\begin{equation}
    (\nabla^2 + \omega^2)\Phi(\vec{r}) = 4 \omega^2U\Phi(\vec{r})~.
\label{eq:waveeqn}
\end{equation}
%%%%%
We can now define an amplification factor in the frequency domain, $F$, as the ratio of the solutions to Eq.~\eqref{eq:waveeqn} for non-zero and zero $U$, with the magnification given by 
%%%%
\begin{equation}
    \mu(\omega, \vec{y}) = |F(\omega, \vec{y})|^2
    \label{eq:muvFsq}
\end{equation}
%%%%%
for a source angular position $\vec{y}$ as defined below Eq.~\eqref{eq:rE}.
Switching variables to $w$ as defined in Eq.~\eqref{eq:defn:w}, we have~\cite{Nakamura:1999:waveoptix}
%%%%
\begin{equation}
    F(w, \vec{y}) = \frac{w}{2\pi i}\int d^2\vec{\rho}e^{i w T(\vec{\rho}, \vec{y})}~,
\end{equation}
%%%%
where $T(\vec{\rho}, \vec{y})$ is the time delay function
%%%%
\begin{equation}
    T(\vec{\rho}, \vec{y}) = \frac{1}{2}|\vec{\rho} - \vec{y}|^2 - \psi(\vec{\rho})
\end{equation}
%%%%%
for a dimensionless lensing potential $\psi(\vec{\rho})$, with $\vec \rho~\equiv~(\DLens/\rE) \vec\theta$. 
For a spherically symmetric lens, $\psi(\vec{\rho})$ only depends on $\rho \equiv |\vec{\rho}|$.
Taking $\theta$ as the angle between $\vec\rho$ and $\vec{y}$, we now have
%%%%%
\begin{equation}
    F(w, y) = - i w e^{i w y^2/2} \int_0^{\infty} d\rho J_0(w \rho y) \rho e^{i w(\frac{1}{2} \rho^2 - \psi(\rho))}~,
    \label{eq:Fvwygen}
\end{equation}
%%%%%%
where
%%%%
\beq
J_0 (z) = \frac{1}{\pi} \int_0^\pi d\theta e^{iz\cos\theta} 
\eeq
%%%%%
is the Bessel function of the first kind of zeroth order. 

The lensing potential $\psi(\rho)$ is obtained by solving the convergence equation~\cite{NarayanBartelmann}
%%%%
\beq
\vec{\nabla}^2_\rho \psi = 2~\frac{\Sigma(\vec{\rho})}{\Sigma_{\rm crit}}~,
\eeq
%%%%
where $\Sigma$ is a surface density obtained by projecting the lens' density onto the lens plane and $\Sigma_{\rm crit} \equiv c^2\DSource/(4\pi G \DLens \DLensSource)$ is a critical mass density. 

{\bf \em Point-like lenses.}
For lenses with extent $\ll \rE$ such as PBHs, $\psi(\rho) = \log \rho$,
so that from Eqs.~\eqref{eq:muvFsq} and \eqref{eq:Fvwygen} and a bit of algebra, we have
%%%%%
\begin{equation}
    \mu(w,y) = \frac{\pi w}{1 - e^{-\pi w}} \bigg|{}_1F_1\bigg(\frac{i w}{2}, 1; \frac{i w y^2}{2}\bigg)\bigg|^2~,
    \label{eq-58mag}
\end{equation}
%%%%%
where we have used the relations
%%%%%
\begin{equation*}
\begin{split}
    \int_0^{\infty} x^m e^{-\alpha x^2} J_n(\beta x) dx = \frac{\beta^n \Gamma(\frac{m+n+1}{2})}{2^{n+1} \alpha^{\frac{m+n+1}{2}} \Gamma(n+1)} \\ \times {}_1F_1\bigg(\frac{m+n+1}{2}, n + 1 ; -\frac{\beta^2}{4\alpha}\bigg)~,
    \label{eq-56mag}
\end{split}
\end{equation*}
%%%
and
%%%%
\begin{equation*}
    \Gamma(1 + i y) \Gamma(1 - i y) = \frac{\pi y}{\sinh(\pi y)}~.
\end{equation*}
%%%%%
In the limit $y \rightarrow 0$, ${}_1F_1 \rightarrow 1$ and we get the maximum magnification during transit as
%%%%
\begin{equation*}
    \mu_{\rm max} = \frac{\pi w}{1 - e^{-\pi w}}~.
\end{equation*}
%%%%%

{\bf \em The effect of finite source size.}
To account for the finite extent of the source, consider a Gaussian distribution of source intensity,
%%%%
\begin{equation}
    W(\vec{y},\vec{r}) = \exp\bigg(-\frac{|\vec{y} - \vec{r}|^2}{2\aSource^2}\bigg)~.
\end{equation}
%%%%
Then the intensity-averaged magnification 
\begin{equation}
    \mu(w,y,\aSource) = \frac{\int d^2 r W(y, r) \mu(w, r)}{\int d^2 r W(y, r)}
\end{equation}
gives Eq.~\eqref{eq:muwaveFSE} under the assumption of circular symmetry, i.e., that $W$ depends only on $r \equiv |\vec{r}|$

%%%%5
\section{Other probes of the PBH mass window}
\label{app:otherwindowfasteners}
%%%%%

We enumerate here alternative proposals in the literature to close the PBH mass window. 

[i] Interference fringes of the two microlensing images showing up as modulations in the energy spectrum, a.k.a. ``femtolensing", with about 100 or more gamma ray burst (GRB) sources in order to overcome the finite source effect~\cite{Katz:2018zrn}.

[ii] Differences in magnification as seen simultaneously by two observing instruments separated by a finite distance, a.k.a. ``parallax microlensing"~\cite{Parallaxmicrolens1995,*Parallaxmicrolens1998}.
With more than 1000 GRBs, the future {\em Daksha} system can probe the PBH mass window with one satellite each orbiting the Earth and the Moon~\cite{Gawade:2023gmt}.

[iii] Transits of light PBHs through the Solar System, estimated to occur about once in a decade, may affect precision ephemerides~\cite{Tran:2023jci} or may be picked up in the future gravitational wave detection interferometer LISA~\cite{Adams:2004pk}.

[iv] Stellar capture of PBHs in dark matter-rich regions such as ultra-faint dwarf galaxies, followed by the transmutation of the host star into a black hole, suppresses the main sequence population in such regions~\cite{EsserTinyakovUltrafaintGal:2022owk,EsserTinyakovUltrafaintGal:2023yut}. 

[v] Neutron star capture of PBHs followed by transmutation of the star into a black hole~\cite{PBHNScaptureCapelaTinyakov:2013yf}.
However, for this effect to be observable the capture rate must be enhanced by large dark matter densities in nearby regions.
Globular clusters may provide one such setting as suggested in Ref.~\cite{PBHNScaptureCapelaTinyakov:2013yf}, but the dark matter content of these systems is highly uncertain for multiple reasons~\cite{Garani:2023esk}. 
In any case, the PBHs are likely to end up in loosely bound orbits around the neutron star and be disrupted by neighbouring stars in dense stellar environments~\cite{Caiozzo:2024flz}.

[vi] PBH transits of white dwarfs may deposit energy via dynamical friction and trigger Type Ia-like supernovae~\cite{Graham:2015apa}, but the heating of nuclear fuel may not occur faster than its cooling by various mechanisms in sizeable portions of the star, making this an unobservable phenomenon for $f_{\rm PBH} \leq 1$~\cite{Montero-Camacho:2019jte}.

\bibliography{refs}

\end{document}